\documentclass[aps,prb,twocolumn,showpacs,amsmath,amssymb]{revtex4}

\usepackage{graphicx}
\usepackage{dcolumn}
\usepackage{bm}
\begin{document}

\title{Vortex Dynamics in Percolative Superconductors
\\ Containing Fractal Clusters of a Normal Phase}

\author{Yuriy I. Kuzmin}
\email{yurk@mail.ioffe.ru}

\affiliation{Ioffe Physical Technical Institute of the Russian
Academy of Sciences, 26 Polytechnicheskaya Street, Saint
Petersburg 194021 Russia}

\date{\today}

\begin{abstract}
The effect of fractal clusters on magnetic and transport
properties of percolative superconductors is studied. The
superconductor contains percolative superconducting cluster
carrying a transport current and clusters of a normal phase. It is
found that normal phase clusters have essential fractal features.
The fractal dimension of the boundary of normal phase clusters is
estimated. The current-voltage (V-I) characteristics of
superconductors containing fractal clusters are obtained. It is
found that the fractality of the cluster boundary intensifies
pinning. This feature permits to enhance the current-carrying
capability of the superconductors.
\end{abstract}


\maketitle

\section{Introduction}

An essential feature of the clusters of defects in superconductor
lies in their capability to trap a magnetic flux.
\cite{indenbom,tonomura,mezzetti,beasley,tpl2000} The cluster
structure affects the vortex dynamics in superconductors,
especially when clusters have fractal boundary.
\cite{surdeanu,prester,pla2000,prb,pla2001,pla2002} A prototype of
superconductor containing inclusion of a normal phase is a
superconducting wire.

The first generation high-temperature superconductor (HTS) wires
are fabricated following the powder-in-tube technique (PIT). The
resulting product consists of one or more superconducting cores
armored by the normal metal. The sheath endows the wire with the
necessary mechanical and electrical properties. The best results
are obtained now for the silver-sheathed bismuth-based composites.
\cite{fukumoto,suenaga} In view of the PIT peculiarity the first
generation HTS wire has a highly inhomogeneous structure.
Superconducting core represents a dense conglomeration of BSCCO
micro-crystallites containing normal phase inclusions inside.

The second generation HTS wires have multi-layered structure
consisting of the metal substrate (nickel-tungsten alloy), the
buffer oxide sub-layer, HTS layer (YBCO), and the protective
cladding made from the noble metal (silver). In the
superconducting layer there are clusters of columnar defects that
can be created during the film growth process as well as by heavy
ion bombardment. Such defects are similar in topology to the
vortices, therefore they suppress effectively the flux creep that
allow to get the critical current up to the depairing value.
\cite{indenbom,tonomura,mezzetti}

\section{Vortex Dynamics in Percolative Superconductors}

A superconductor containing isolated clusters of a normal phase
allows for effective pinning, because the vortices cannot leave
them without crossing the superconducting space. The clusters
consist of sets of normal phase inclusions, united by the common
trapped flux and surrounded by the superconducting phase.
\cite{pla2000,prb} When the current is increased the vortices
start to break away from the clusters of pinning force weaker than
the Lorentz force. Then, the vortices will pass through the weak
links, connecting the normal phase clusters. In this case
depinning has percolative character, \cite{yamafuji,ziese} because
vortices move through randomly generated channels. Weak links form
readily in HTS's due to the intrinsically short coherence length.
\cite{sonier}

Let us take the area of cluster cross-section by the plane
carrying the transport current as a measure of its size. Magnetic
flux trapped into a
cluster is proportional to its area. Hence the decrease in the trapped flux $%
\Delta $$\Phi $ can be expressed with the probability $W\left(
A\right) =\Pr \left\{ \forall A_{j}<A\right\} $ of finding the
clusters of area $A_{j}$
smaller than a given magnitude $A$:%
\begin{equation}
\frac{\Delta \Phi }{\Phi }=1-W(A)  \label{prob1}
\end{equation}

\hspace{5mm} According to weak link configuration each normal
phase cluster has its own value of depinning current, which
contributes to the overall distribution. Thus the decrease in the
trapped flux is proportional to the number of all the normal phase
clusters of critical currents less than a preset value $I$ and can
be expressed with the probability $F\left( I\right)
=\Pr \left\{ \forall I_{j}<I\right\} $ :%
\begin{equation}
\frac{\Delta \Phi }{\Phi }=F\left( I\right)   \label{probi2}
\end{equation}

The critical current and cluster area distributions are
interdependent, because larger cluster has more weak links over
its boundary and, consequently, the smaller depinning current.

\section{Weak Link Distribution over the Cluster Boundary}

Let us analyze how the vortex exits from a normal phase cluster.
As the transport current increases, the Lorentz force, pushing the
vortex out, will increase as well. In order to leave the cluster
the vortex has to enter into one of the weak links, which are
randomly arranged along the cluster perimeter. Suppose that after
the vortex reaches the entry point it passes all the way between
two adjacent normal phase clusters without being trapped inside
the weak link itself. The vortex exit can be considered as the
result of random walks under the action of the Lorentz force. The
following outcomes of the random walks may happen: (a) the vortex
enters the weak link and leaves the normal phase cluster, (b) the
vortex does not enter the weak link and continues its random
walks, and (c) the vortex does not enter the weak link at all and
remains to be locked inside. The mean number of the entry points
on the cluster perimeter gives the probability measure of the
random walk outcomes, which are favorable for the vortex to go
out. So an exit of the vortex from a normal cluster can be treated
as the two-dimensional generalization of the problem of a random
walk particle reaching an absorbing border. Unlike the classic
problem of the distribution of the exit points, \cite{spitzer}
here the boundary of the area is not absorbing all over, but there
are only discrete absorption points where the vortices can enter
the weak links. Moreover, the situation is complicated by the fact
that a random walker is permanently subjected to the Lorentz
force.

The distribution of entry points over cluster perimeter varies
from one cluster to another, so that each cluster has the entry
point distribution function $\psi \left( l\right) $ of its own,
where $l$ is the co-ordinate along perimeter. The probability
distribution of functions $\psi \left(
l\right) $ over all the clusters can be characterized by the probability Pr$%
\{\psi \left( l\right) \}$ of finding a given function $\psi
\left( l\right) $.

The most probable function of entry point distribution is the mean
over all functions

\begin{equation}
\overline{\psi \left( l\right) }=\int\limits_{(\Omega )}D\psi
\left( l\right) \psi \left( l\right) \Pr \left\{ \psi \left(
l\right) \right\} \label{path3}
\end{equation}

The path integral Fourier transform on the probability functional
Pr$\{\psi \left( l\right) \}$ represents the characteristic
functional
\begin{equation}
H\left[ k(l)\right] =\frac{\int\limits_{\left( \Omega \right) }\mathcal{D}%
\psi \left( l\right) \,\exp \left( i\oint dl\,k\left( l\right)
\psi \left(
l\right) \right) \Pr \left\{ \psi \left( l\right) \right\} }{%
\int\limits_{\left( \Omega \right) }\mathcal{D}\psi \left(
l\right) \,\Pr \left\{ \psi \left( l\right) \right\} }
\label{fourier4}
\end{equation}%
where $k=k\left( l\right) $ is the element of a reciprocal
function space.

If all the clusters are of an equal entry point distribution,
which coincides with the most probable one (\ref{path3}), the
probability $\Pr \left\{ \psi \left( l\right) \right\} $ is zero
for all $\psi \left( l\right) $ that differ from $\overline{\psi
\left( l\right) }$, whereas $\Pr \left\{ \overline{\psi \left(
l\right) }\right\} =1$ . At that rate the functional
(\ref{fourier4}) becomes

\begin{equation}
H\left[ k\left( l\right) \right] =\exp \left( i\oint dlk\left(
l\right) \overline{\psi \left( l\right) }\right)   \label{char5}
\end{equation}

If all the entry points are uniformly distributed over the cluster
perimeter, the functional takes the form

\begin{equation}
H\left[ k\left( l\right) \right] =\exp \left( i\frac{\beta
N}{P}\oint dl\,k\left( l\right) \right)   \label{charf6}
\end{equation}%
where the constant $\beta $ is being chosen to normalize the
distribution function $\psi \left( l\right) $ to unity, so that
$\beta N=1$. The functional (\ref{charf6}) has the form of
equation (\ref{char5}) for the
uniform distribution of entry points: $\overline{\psi \left( l\right) }=1/P$%
. This means that all the clusters have the same uniform
distribution of the entry points, for which the probability of
finding a weak link at any point of the perimeter is independent
of its position.

If the concentration of entry points per unit perimeter length $n=\overline{N%
}/P$ is constant for all clusters, and all the clusters are
statistically self-similar, the mean number of entry points
$\overline{N}$ along perimeter is proportional to its length
$\overline{N}=\oint n(l)dl=nP$ . The more entry points into weak
links are accessible for random walk vortices driven by the
Lorentz force, the more is the probability that the vortex will
leave the cluster, and therefore, the smaller is the Lorentz force
required to push the vortex out. Hence, we may write the following
relationship between the critical current of the cluster, at which
the magnetic flux ceases to be trapped inside, and its geometric
size $I\propto 1/\overline{N}\propto 1/P$ .

Thus, to deal with the distribution function (\ref{prob1}), the
relation between perimeter and area of clusters should be studied.
It might be natural to suppose that the perimeter-area relation
obeys the well known geometric formula $P\propto \sqrt{A}$.
However, it would be a very rough approximation, because this
relationship holds for Euclidean geometric objects only. As it was
first found in, \cite{pla2000} the normal phase clusters can have
fractal boundaries, i. e. the perimeter of their cross-section and
the enclosed area obey the scaling law

\begin{equation}
P^{\frac{1}{D}}\propto A^{\frac{1}{2}}  \label{scaling7}
\end{equation}%
where $D$ is the fractal dimension of the cluster boundary.
\cite{mandelbrot} The fractal nature of such clusters affects the
vortex transport and depinning in superconductors. \cite{prb}

\section{Fractal Geometry of Normal Phase Clusters}

In order to clear up how the developed approach can be used in
practice, the electron photomicrographs of YBCO films have been
studied. The films were prepared by magnetron sputtering on
sapphire substrates with a cerium oxide buffer sublayer. The
normal phase clusters were formed by columnar inclusions of
nonstoichiometric composition. These inclusions were created at
the sites of defects on the boundary with the substrate in such a
way that they were oriented normally to the film surface. The
profiles of cluster sections by the film plane were clearly
visible on the photomicrographs, and their perimeters and enclosed
areas have been measured. So the cross-sections of the extended
columnar objects, which the normal phase clusters are, have been
investigated. The normal phase has occupied 20\% of the total
surface so the percolative superconducting cluster was dense
enough. The perimeters and areas of clusters have been
measured by covering their digitized pictures with a grid of spacing 60$%
\times $60nm$^{2}$. The sampling has contained 528 normal phase
clusters located on the scanned area of 200$\mu $m$^{2}$. A high
skewness (1.765) as well as the statistically insignificant (5\%)
difference between the mean cluster area and the standard
deviation has confirmed that the distribution of the cluster areas
is exponential.

\begin{figure}
\includegraphics{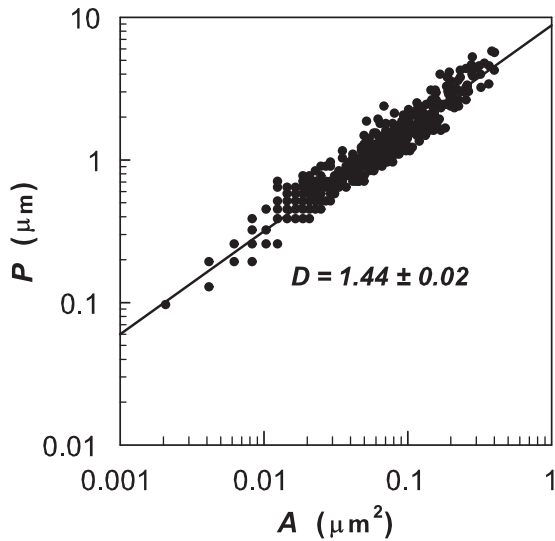}
\caption{\label{fig1} Perimeter-area relationship for the normal
phase clusters with fractal boundary in YBCO film. The solid line
indicates the least square regression line.}
\end{figure}

The obtained data are given in Fig.~\ref{fig1}. All the points
fall on a straight line on double logarithmic scale with
correlation coefficient of 0.929. This graph shows that the
perimeter-area scaling relation, which is inherent to fractals, is
valid in the range of almost three orders of magnitude in cluster
area. The slope of the perimeter-area regression line gives the
fractal dimension of the cluster boundary $D=1.44\pm 0.02$. This
fact that the scaling law (\ref{scaling7}) with fractional
exponent $D$ is fulfilled for the normal phase clusters, gives an
evidence for their fractal nature.

In the general way the cluster area statistics may be described by
gamma distribution

\begin{equation}
W\left( A\right) =\left( \Gamma \left( g+1\right) \right)
^{-1}\gamma \left( g+1,\frac{A}{A_{0}}\right)   \label{cuma8}
\end{equation}%
where $\Gamma \left( \nu \right) $ is Euler gamma function,
$\gamma \left( \nu \right) $ is the incomplete gamma function,
$A_{0}$ and $g$ are the
parameters of gamma distribution that control the mean area of the cluster $%
\overline{A}=\left( g+1\right) A_{0}$ and its variance $\sigma
_{A}^{2}=\left( g+1\right) A_{0}^{2}$. Exponential distribution is
the simplest case of gamma distribution of $g=0$.

In accordance with starting formulas (\ref{prob1}), (\ref{probi2})
as well as with scaling law (\ref{scaling7}) we can get the
relation between the critical current of the cluster and its
geometric size: $I=\alpha A^{-D/2}$,
where $\alpha $ is the cluster form factor. The cluster area distribution (%
\ref{cuma8}) gives rise to the critical current distribution

\begin{equation}
F\left( i\right) =\left( \Gamma \left( g+1\right) \right)
^{-1}\Gamma \left( g+1,Gi^{-2/D}\right)   \label{cumi9}
\end{equation}%
where $G\equiv \left( \theta ^{\theta }/\left( \theta
^{g+1}-\left( D/2\right) \exp \left( \theta \right) \Gamma \left(
g+1,\theta \right) \right) \right) ^{2/D}$, $\theta \equiv
g+1+D/2$,\ $\Gamma \left( \nu ,z\right) $ is the complementary
incomplete gamma function, $i=I/I_{c}$ is the dimensionless
electric current, $I_{c}=\alpha \left( A_{0}G\right) ^{-D/2}$ is
the critical current, which gives the point of intersection of the
current axis and the tangent line drawn through the inflection
point of the dependence of differential resistance on the current.
The found
distribution (\ref{cumi9}) allows to derive the probability density $%
f(i)\equiv dF/di$ for the critical currents

\begin{equation}
f(i)=\frac{2G^{g+1}}{D\Gamma (g+1)}i^{-\left( 2/D\right)
(g+1)-1}\exp \left( -Gi^{-2/D}\right)   \label{densi10}
\end{equation}

\section{Resistive State of Superconductors with Fractal Clusters of a
Normal Phase}

Each normal phase cluster contributes to the total critical
current distribution, so the voltage across a sample can be
represented as the response to the sum of effects from each
cluster

\begin{equation}
\frac{V}{R_{f}}=\int\limits_{0}^{i}\left( i-i^{\prime }\right)
f\left( i^{\prime }\right) di^{\prime }  \label{conv11}
\end{equation}%
where $R_{f}$ is the flux flow resistance. Thus, using the
critical current distribution (\ref{densi10}), we get the V-I
characteristics:

\begin{eqnarray}
\frac{V}{R_{f}} &=&\frac{1}{\Gamma \left( g+1\right) }\Biggl(
i\Gamma \left(
g+1,Gi^{-2/D}\right)   \notag \\
&&-G^{D/2}\Gamma \left( g+1-\frac{D}{2},Gi^{-2/D}\right) \Biggr)
\label{gamma12}
\end{eqnarray}

\begin{figure}
\includegraphics{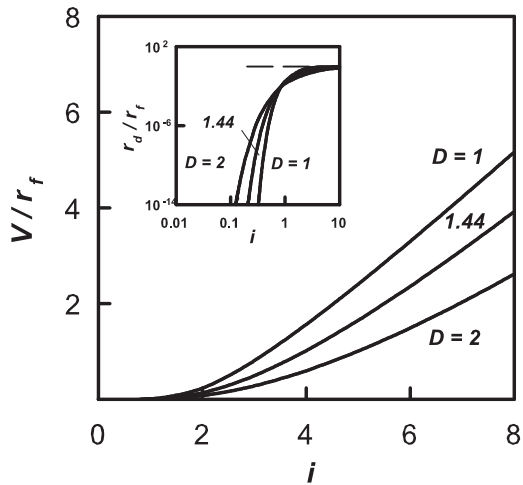}
\caption{\label{fig2} Current-voltage characteristics of
superconductor containing fractal clusters of a normal phase. The
inset shows the dependence of differential resistance on a
transport current, which is typical for the vortex glass.}
\end{figure}

The V-I curves for the different fractal dimensions are presented
in Fig.~\ref{fig2}. The lines drawn for Euclidean clusters ($D=1$)
and for the clusters of the most fractality ($D=2$) bound the
region the V-I characteristics can fall within. The curve between
them gives the case of the previously found fractal dimension
$D=1.44$. The inset in Fig.~\ref{fig2} shows the dependence of
differential resistance $r_{d}\equiv dV/di$ on the transport
current. Differential resistance is proportional to density of
vortices $n$ broken away from the pinning centers:
$r_{d}=R_{f}n\Phi _{0}/B$, where $B$ is the magnetic field, $\Phi
$$_{0}$$\equiv hc/(2e)$ is the magnetic flux quantum, $h$ is
Planck constant, $c$ is the velocity of light, and $e$ is the
electron charge. It is just a motion of free vortices induces
electrical field. The dependencies of resistance on the current
shown in this graph are typical for the vortex glass: the curves
have a convex form on double logarithmic scale as well as
resistance tends to zero with decreasing current as a result of
flux creep suppression. \cite{blatter,brown}

\begin{figure}
\includegraphics{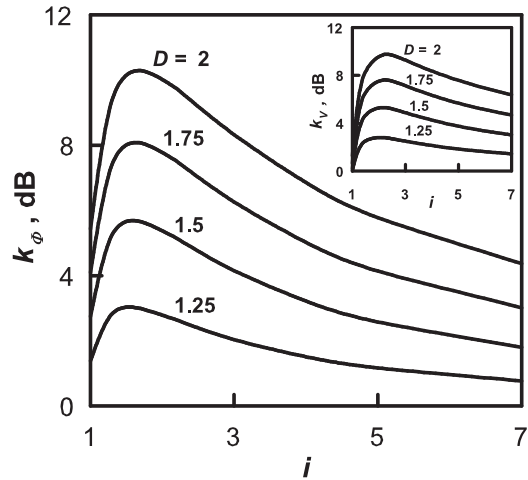}
\caption{\label{fig3} Pinning gain for the different fractal
dimensions of the cluster boundary. In the inset the voltage
attenuation factor is shown.}
\end{figure}

All the V-I characteristics in Fig.~\ref{fig2} are virtually
starting with the point of $i=1$. It is seen that the cluster
fractality reduces electric field arising from the vortex motion.
The reason of this
phenomenon lies in the peculiarity of the critical current distribution (\ref%
{densi10}). As the fractal dimension increases, this distribution
broadens out, moving towards greater magnitudes of current. It
means that more and more of the small clusters, which can trap the
vortices best, are being involved in the process. The smaller part
of the vortices can move, the weaker the induced electric field.
Resistive transition widens, shifting towards higher currents. The
effect of fractal dimension can be characterized by the pinning
gain factor $k_{\Phi }\equiv 20\log \left( \Delta \Phi \left(
D=1\right) /\Delta \Phi \left( D\right) \right) $, which is equal
to relative decrease in the number of vortices broken away from
the clusters of fractal dimension \textit{D} compared to the
Euclidean ones, as well as by the voltage attenuation factor
$k_{V}\equiv 20\log \left( V\left( D\right) /V\left( D=1\right)
\right) $, characterizing reduction of electric field. The
dependencies of these factors on the transport current for
different fractal dimensions are shown in Fig.~\ref{fig3}. It is
necessary to keep in mind that these curves, as well as the graphs
inFig.~\ref{fig2}, are plotted versus dimensionless current
normalized relatively to the critical current $I_{c}$, the value
of which depends on the fractal dimension. The pinning gain
characterizes the properties of a superconductor in the range of
the transport currents $i>1$. At smaller current the breaking of
the vortices away has not started yet for lack of pinning centers
of such small critical currents. The pinning enhancement due to
the cluster fractality in the neighborhood of resistive transition
can only be realized in the case of efficient heat removal that
prevents the development of thermo-magnetic instability. As for
any hard superconductor the energy dissipation in the resistive
state does not mean the destruction of phase coherence yet. Some
dissipation always accompanies any motion of vortices that can
happen in even at low transport current. Superconducting state
collapses only when a growth of dissipation becomes avalanche-like
as a result of thermo-magnetic instability.

\section{Conclusion}

In the present work the fractal nature of the normal phase
clusters is revealed. It is found that the fractality of cluster
boundary strengthens the flux pinning. This feature gives the new
possibility for increasing the current-carrying capability of
composite superconductors by optimization the material structure
without changing of its chemical composition.

\bibliography{ASC}

\end{document}